\documentclass[aps,prb,reprint,superscriptaddress,preprintnumbers,showpacs,amsmath,amssymb]{revtex4-1}
\usepackage{bm}
\usepackage[colorlinks=true,linkcolor=blue,citecolor=blue]{hyperref}
\usepackage{times}
\usepackage{physics}
\usepackage{amsmath}
\usepackage{amssymb}
\usepackage{amsthm}
\usepackage{amsfonts}
\usepackage{enumerate}
\usepackage{latexsym}
\usepackage{ifpdf}
\usepackage{graphicx}
\usepackage{makeidx}
\expandafter\ifx\csname package@font\endcsname\relax\else
 \expandafter\expandafter
 \expandafter\usepackage
 \expandafter\expandafter
 \expandafter{\csname package@font\endcsname}%
 \fi
\usepackage{color}

\begin{document}

\title{Synthesis and electronic properties of Ruddlesden-Popper strontium iridate epitaxial thin films stabilized by control of growth kinetics}

\author{Xiaoran Liu}
\email{xiaoran.liu@rutgers.edu}
\affiliation{Department of Physics and Astronomy, Rutgers University, Piscataway, New Jersey 08854, USA}

\author{Yanwei Cao}
\affiliation{Department of Physics and Astronomy, Rutgers University, Piscataway, New Jersey 08854, USA}
\affiliation{Ningbo Institute of Materials Technology and Engineering, Chinese Academy of Sciences, Ningbo, Zhejiang 315201, China}

\author{B. Pal}
\affiliation{Department of Physics and Astronomy, Rutgers University, Piscataway, New Jersey 08854, USA}

\author{S. Middey}
\affiliation{Department of Physics, Indian Institute of Science, Bengaluru 560012, India}

\author{M. Kareev}
\affiliation{Department of Physics and Astronomy, Rutgers University, Piscataway, New Jersey 08854, USA}

\author{Y. Choi}
\affiliation{Advanced Photon Source, Argonne National Laboratory, Argonne, Illinois 60439, USA}

\author{P. Shafer}
\affiliation{Advanced Light Source, Lawrence Berkeley National Laboratory, Berkeley, California 94720, USA}

\author{D. Haskel}
\affiliation{Advanced Photon Source, Argonne National Laboratory, Argonne, Illinois 60439, USA}

\author{E. Arenholz}
\affiliation{Advanced Light Source, Lawrence Berkeley National Laboratory, Berkeley, California 94720, USA}

\author{J. Chakhalian}
\affiliation{Department of Physics and Astronomy, Rutgers University, Piscataway, New Jersey 08854, USA}

\begin{abstract}

We report on the selective fabrication of high-quality Sr$_2$IrO$_4$ and SrIrO$_3$ epitaxial thin films from a single polycrystalline Sr$_2$IrO$_4$ target by pulsed laser deposition. Using a combination of X-ray diffraction and photoemission spectroscopy characterizations, we discover that within a relatively narrow range of substrate temperature, the oxygen partial pressure plays a critical role in the cation stoichiometric ratio of the films, and triggers the stabilization of different Ruddlesden-Popper (RP) phases. Resonant X-ray absorption spectroscopy measurements taken at the Ir $L$-edge and the O $K$-edge demonstrate the presence of strong spin-orbit coupling, and reveal the electronic and orbital structures of both compounds. These results suggest that in addition to the conventional thermodynamics consideration, higher members of the Sr$_{n+1}$Ir$_n$O$_{3n+1}$ series can possibly be achieved by kinetic control away from the thermodynamic limit. These findings offer a new approach to the synthesis of ultra-thin films of the RP series of iridates and can be extended to other complex oxides with layered structure.     

\end{abstract}

\maketitle

\newpage

\section{Introduction}

In recent years, 5$d$ transition metal oxides have attracted a tremendous research interest due to the comparable strength of the on-site electron-electron correlations $U$ and the spin-orbit coupling (SOC) $\lambda$, which can give rise to a plethora of exotic quantum states of matter including topological insulator, quantum spin liquid, Weyl semimetals, spin-orbit Mott insulator, etc.\cite{Krempa_ARCMP_2014,Schaffer_RPP_2016} A prototypical example is the Ruddlesden-Popper (RP) series (Sr$_{n+1}$Ir$_n$O$_{3n+1}, n$ = 1, 2, $\ldots$, $\infty$) of iridium oxides (Ir$^{4+}$, 5$d^5$), where the electronic structure exhibits distinct variations as a function of $n$, which is effectively controlled by the dimensionality of these compounds.\cite{Moon_PRL_2008} On one hand, in layered perovskite Sr$_2$IrO$_4$ ($n$ = 1), the t$_{2g}$ band is split by the strong SOC, leading to the formation of $J_\textrm{eff} = 1/2$ and $J_\textrm{eff} = 3/2$ subbands. A modest $U$ further opens a gap and splits the narrow $J_\textrm{eff} = 1/2$ band into upper Hubbard band (UHB) and lower Hubbard band (LHB), giving rise to a unique spin-orbit entangled Mott insulating ground state with antiferromagnetic long range ordering.\cite{Kim_PRL_2008, Kim_Science_2009} Moreover, in view of its close similarity to cuprates in both the structure and the electronic behaviors \cite{Kim_PRL_2012, Kim_NPhys_2015,Zhao_NPhys_2015,Yan_PRX_2015}, doped Sr$_2$IrO$_4$ has been regarded as a promising platform to realize high-temperature superconducting iridates.\cite{Wang_PRL_2011,Watanabe_PRL_2013,Meng_PRL_2014} On the other hand, for the perovskite SrIrO$_3$ ($n$ = $\infty$), the increased Ir 5$d$ bandwidth $W$, together with  comparable  in the magnitude $\lambda$ and $U$, eventually prevents a Mott gap opening and results in an intriguing correlated semi-metallic ground state \cite{Moon_PRL_2008, Zeb_PRB_2012}; Based on this consideration theorists have predicted the formation of artificial topological insulating phases.\cite{Xiao_NC_2011,Carter_PRB_2012, Chen_NC_2015}       

Lately, to exploit the possibility of novel emergent phenomena caused by the pronounced SOC effect, active fabrication efforts for Sr$_2$IrO$_4$ (Sr214) and SrIrO$_3$ (Sr113) epitaxial thin films, superlattices, and heterostructures have been put forward  \cite{Jian_arkiv_2013,Lee_PRB_2012, Jian_PRB_2013,Nichlos_APL_2013, Nichlos_APL_2013_2, Miao_PRB_2014, Biswas_JAP_2014, Zhang_PRB_2015, Nie_PRL_2015, Liu_Srep_2016,Fina_NC_2013, Matsuno_PRL_2015, Groenendijk_APL_2016,Hirai_APLMater_2015,Yi_PNAS_2015, Nichols_NC_2016,Nishio_APL Mater_2016, Seo_APL_2016, Gruenewald_AM_2017, Hao_PRL_2017, Okamoto_NanoLett_2017}. Among those, the majority of the films were synthesized by pulsed laser deposition (PLD) in which the proper stoichiometry of the film is typically achieved from a chemically equivalent target. Very recently, however, it was noticed that off-stoichiometry can occur when using either SrIrO$_3$ and Sr$_2$IrO$_4$ target as the source \cite{Nishio_APL Mater_2016, Seo_APL_2016}. Due to the highly non-equilibrium nature of the ablation process, the kinetics of the RP Sr$_{n+1}$Ir$_n$O$_{3n+1}$ epitaxial growth such as plume propagation or crystallization at substrate surface has been thus far little explored.
                      
In this paper, we demonstrate that in addition to the conventional thermodynamic considerations, a selective fabrication of different Sr$_{n+1}$Ir$_n$O$_{3n+1}$ ($n = 1, \infty$) epitaxial films can be achieved from a single Sr214 target by virtue of the kinetic nature of PLD. A combination of X-ray diffraction and d.c. transport measurements confirm the formation of both proper chemical composition and excellent structural quality of the samples. By systematically varying the oxygen partial partial P$_{O_2}$ and the substrate temperature $T_s$, we find that the oxygen partial pressure plays a decisive role in the cation stoichiometric ratio of the film leading to the stabilization of various RP phases, and the proper substrate temperature mainly enables the persistent two-dimensional (2D) growth mode. Synchrotron based X-ray absorption measurements at Ir $L$-edge indicate large expectation values of the spin-orbit coupling for both compounds. Polarization-dependent absorption spectra taken at O $K$-edge reveal the presence of an expected strong hybridization between Ir 5$d$ and O 2$p$ states within each film.

\section{Experiments}

The basic unit cell of Sr214 and Sr113 are shown in Fig.~1. As seen, for the starting member Sr214 ($n = 1$), where integer $n$ refers to the number of the perovskite blocks sandwiched between the extra rock-salt SrO layers, the network of the corner-sharing IrO$_6$ octahedra persist two dimensionally in the $ab$ plane and is essentially disrupted along the $c$ axis. The tilts and rotations of the IrO$_6$ octahedra further expand the unit cell by $\sqrt{2} \times \sqrt{2} \times 2$ along three axes, leading to the tetragonal $I4_1/acd$ crystal structure. In contrast, for the end-member Sr113 ($n = \infty$)  the corner-sharing IrO$_6$ octahedra are connected three dimensionally with tilts and rotations along each axis, giving rise to a perovskite structure with orthorhombic $Pbnm$ space group. It is important to note that under ambient pressure Sr113 crystalizes into a monoclinic 6M rather than perovskite structure.\cite{Longo_JSSC_1971} Therefore high pressure environment or epitaxial compressive strain is needed for stabilizing the proper perovskite phase.   

The strontium iridate Sr$_{n+1}$Ir$_n$O$_{3n+1}$  ($n = 1, \infty$) thin films were grown on (001) SrTiO$_3$ (STO) substrates, by varying the oxygen partial pressure $P_{O_2}$, and the substrate temperature $T_{s}$, from 10$^{-6}$ to 10$^{-1}$ Torr, and from 550$^{\circ}$C to 700 $^{\circ}$C, respectively. The detailed lattice parameters of bulk Sr214, Sr113, and STO are given in Table I. During the growth, a single stoichiometric Sr214 polycrystalline target was used as the ablation source for all reported samples. The fluence and the repetition rate of the KrF excimer laser ($\lambda=248$ nm) were fixed at about 2 J/cm$^2$ and 2 Hz, respectively. The substrate-to-target distance was set at 60 mm. To explore the growth kinetics the entire deposition process was monitored by {\it{in-situ}} high-pressure reflection-high-energy-electron-diffraction (HP RHEED). After growth, samples were cooled down to room temperature at a rate of 15 $^{\circ}$C/min.

\begin{figure}[t]\vspace{-0pt}
\center
\includegraphics[width=0.5\textwidth]{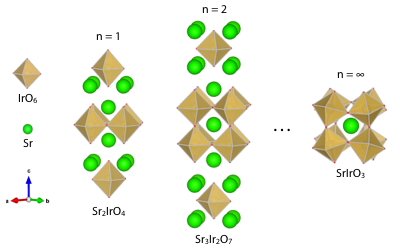}
\caption{\label{} Crystal structures of Sr$_{n+1}$Ir$_n$O$_{3n+1}$ Ruddlesden-Popper phases: Sr$_2$IrO$_4$ (n = 1), Sr$_3$Ir$_2$O$_7$ (n = 2), and SrIrO$_3$ (n = $\infty$).}
\end{figure}

The chemical composition of the films was investigated by X-ray photoelectron spectroscopy (XPS). The measurements were carried out in a Thermo Scientific X-ray Photoelectron Spectrometer System equipped with a hemispherical analyzer and a monochromatic Al K$_\alpha$ source. The photoelectrons were collected in a surface normal geometry in order to increase the bulk sensitivity. The spectra were calibrated using C 1$s$ spectra with binding energy close to 284.6 eV. Decomposition of each spectrum was performed with the casaXPS software using Gaussian-Lorentz line profiles. Structural properties of the samples were characterized by high resolution X-ray diffraction (XRD) and reciprocal space mapping (RSM) measurements using a PANalytical Material Research Diffractometer (Cu K$_{\alpha1}$ line, $\lambda$ = 1.5406 {\AA}). The d.c. electrical transport properties were performed with a Physical Property Measurement System (PPMS, Quantum Design) in the Van der Pauw geometry. The X-ray absorption spectra (XAS) on Ir $L_{2,3}$ edges were performed at beamline 4-ID-D of the Advanced Photon Source in Argonne National Laboratory. Data were collected using a helicity modulation technique at 10 K and recorded in the fluorescence yield (FY) detection mode. In addition, the O $K$-edge polarization dependent XAS were measured at beamline 4.0.2 of the Advanced Light Source in Lawrence Berkeley National Laboratory. Those data were collected in the total electron yield (TEY) detection mode at 300 K.

\begin{table}[t]
\caption{Lattice parameters of bulk Sr$_2$IrO$_4$, SrIrO$_3$ and SrTiO$_3$. The $a$, $b$ and $c$ represent the values of the conventional unit cell while the $a_c$, $b_c$ and $c_c$ are for the pseudocubic unit cell. Note, $a_c = b_c = \sqrt{a^2 + b^2} / 2$, and $c_c = c/2$. The epitaxial strain, $\varepsilon = (a_{\textrm{substrate}} - a_{\textrm{bulk}})/a_{\textrm{bulk}} \times 100\%$. Compressive strain if $\varepsilon < 0$ and tensile strain if $\varepsilon > 0$.}
\centering
\setlength{\tabcolsep}{4.5pt}
\begin{tabular}{c c c c c c c}
\hline\hline
Material & $a$ (\AA) & $b$ (\AA) & $c$ (\AA) & $a_c = b_c$ (\AA) & $c_c$ (\AA) & $\varepsilon$ (\%) \\ [0.5ex]
\hline
Sr$_2$IrO$_4$ & 5.49 & 5.49 & 25.78 & 3.88 & 12.89 & +0.52\% \\
SrIrO$_3$ & 5.60 & 5.58 & 7.89 & 3.95 & 3.95 & -1.26\% \\
SrTiO$_3$ & 3.90 & 3.90 & 3.90 & \ & \ & \ \\
\hline\hline
\end{tabular}
\label{Table I}
\end{table}

\section{Results and Discussion}

\subsection{Synthesis}

\begin{figure}[t]\vspace{-0pt}
\center
\includegraphics[width=0.5\textwidth]{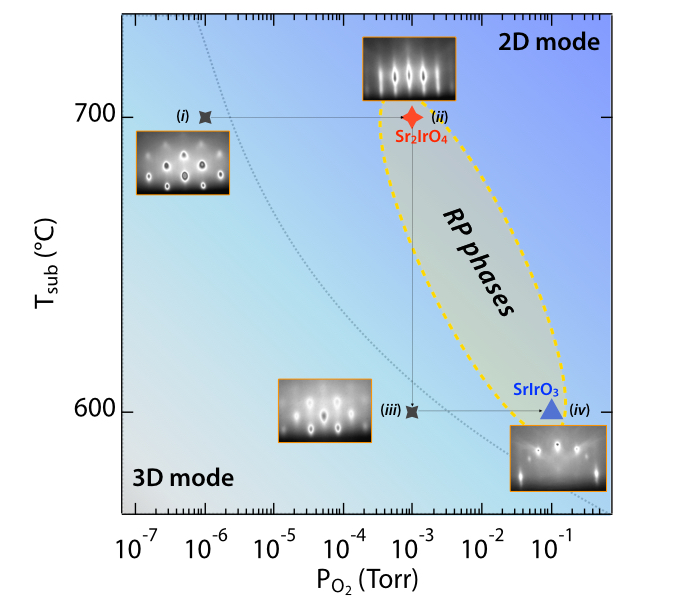}
\caption{\label{} Growth phase diagram of RP Sr$_{n+1}$Ir$_n$O$_{3n+1}$ thin films. The black dotted line represents the estimated boundary between 3D and 2D growth modes. In each case, the {\it{in-situ}} RHEED image recorded after the deposition is exhibited on the figure. The highlighted area indicates the possible growth windows for other RP phases.}
\end{figure}

\begin{figure}[t]\vspace{-0pt}
\center
\includegraphics[width=0.45\textwidth]{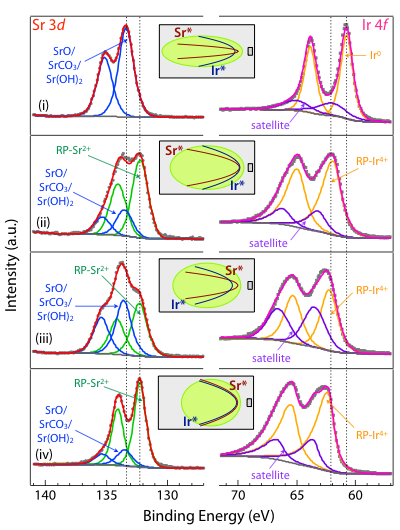}
\caption{\label{} Core level Sr 3$d$ (left) and Ir 4$f$ (right) XPS of samples fabricated at various conditions. Black circles: experimental data; Red and purple lines: summation envelope of Sr and Ir, respectively; Blue line: component from surface SrO/SrCO$_3$/Sr(OH)$_2$, green line: component from RP films; Orange line: component from Ir$^*$ species; Violet line: component from Ir$^*$ satellite; Gray line: Shirley type background. The width of both doublets were kept equal and the intensity ratio of Ir 4$f_{7/2}$ to Ir 4$f_{5/2}$ peak was fixed to 4:3 and that of Sr 3$d_{5/2}$ to Sr 3$d_{3/2}$ peak was fixed to 3:2, respectively. Note, spectra were measured at several different areas across the sample to exclude non-uniformity issue. The corresponding angular distributions of Sr$^*$ and Ir$^*$ species within the plume under each growth condition are depicted on the graph.}
\end{figure}

To obtain microscopic insight into the phase formation, a set of {\it{in-situ}} RHEED and {\it{ex-situ}} XPS measurements was taken on four films grown at different oxygen pressure ($P_{O_2}$) and substrate temperature ($T_{sub}$). As shown in Fig.~2, the growth conditions of the single phase Sr214 [sample ($ii$)] and Sr113 [sample ($iv$)]  were optimized at $P_{O_2} = 10^{-3}$ Torr, $T_{sub} = 700 ^{\circ}$C, and $P_{O_2} = 10^{-1}$ Torr, $T_{sub} = 600 ^{\circ}$C, respectively. The RHEED streaks (specular and off-specular) are clearly seen and well distributed around  the Laue rings from all films with distinct Kikuchi lines, indicative of flat surface and high crystallinity of the samples. The oscillation and recovery of the RHEED intensity further establish the presence of the 2D growth mode for each film. In particular, the two faint spots shown on sample (iv) indicate that the SrIrO$_3$ film possesses the \textit{c-Pbnm} phase \cite{SI}. However, three-dimensional (3D) growth appears for $P_{O_2} = 10^{-6}$ Torr, $T_{sub} = 700 ^{\circ}$C [sample ($i$)], or $P_{O_2} = 10^{-3}$ Torr, $T_{sub} = 600^{\circ}$C [sample ($iii$)], as evidenced by the electron transmission-like RHEED patterns \cite{Ichimiya_RHEED}. 

Next, we investigate the charge states and stoichiometry of the films by XPS. First, since chemical reactions of Sr  with CO$_2$ or H$_2$O on surface of the films can easily take place, the Sr 3$d$ spectra contain contributions from SrO/SrCO$_3$/Sr(OH)$_2$ secondary phases \cite{Mutoro_EES_2011}. Those contributions are found in all samples and expressed by the blue line at $\sim$133.5 eV for 3$d_{5/2}$ and $\sim$135.3 eV for 3$d_{3/2}$. In addition, the component at lower binding energy shown as green line at $\sim$132.3 eV for 3$d_{5/2}$ and $\sim$134.1 eV for 3$d_{3/2}$ is attributed to the  chemically distinct Sr species located beneath the surface, thus reflecting the information from the RP phases \cite{Mutoro_EES_2011}. Note, as shown on Fig.~3 left panel, the Sr 3$d$ spectrum of the sample (i) is well described by single component from SrO/SrCO$_3$/Sr(OH)$_2$, indicating that no RP phase forms at this growth condition and the obtained 3D islands likely constitute of thermodynamically stable SrO or Sr(OH)$^-$ clusters. \cite{Xu_SciRep_2016} In sharp contrast, the spectra of the samples (ii) -- (iv) contain components from both the SrO/SrCO$_3$/Sr(OH)$_2$ and the RP phases. In particular, in the case of the 2D Sr214 (ii) and Sr113 (iv) films, the relative intensity of the impurity peak is rather weak implying that the  thickness of the secondary phases of only a few \AA~at the surface region. However, for sample (iii)\ the relative intensity of the impurity peak becomes much stronger compared to the RP samples, implying a large increase in the formation of SrO clusters on top of the RP layers during the initial stage of deposition. 

Moreover, each of the Ir 4$f$ spectrum exhibits distinctly asymmetric spectral features which have been decomposed in two components, with one relatively intense doublet and one relatively weak doublet, as shown on the right panel of Fig.~3. The intense doublet (orange line) represents the spin-orbit splitting peaks (i.e. Ir 4$f_{7/2}$ and Ir 4$f_{5/2}$).\cite{Yang_SciRep_2017} Interestingly, each Ir 4$f$ spectrum also contains two additional broad peaks (violet line) at higher binding energy with identical spin-orbit strength (peak splitting of $\sim$3.1 eV). The appearance of such additional broad features can be assigned to either the emergence of two different final states, i.e. screened and un-screened core holes \cite{Pfeifer_SIA_2015, Kahk_PRL_2014} or the existence of plasmon satellites in the higher binding energy range \cite{Payne_CPL_2005, Bourlange_JAP_2009}. According to the Kotani model \cite{Kotani_JESRP_1996}, for narrow band metals the core hole generated in the photoemission process interacts with conduction carriers and thereby creates two different (screened and un-screened) final sates. On the other hand, plasmonic satellites always appear at higher binding energy and becomes broadened by conduction electron scattering; In general, the intensity of the plasmonic satellite increases with decreasing  electron density. In our samples, the presence of plasmonic satellite is more likely since intensity of the satellite is significantly higher for insulating Sr214 compared to semi-metallic Sr113.

\begin{figure*}[t]\vspace{-0pt}
\center
\includegraphics[width=0.8\textwidth]{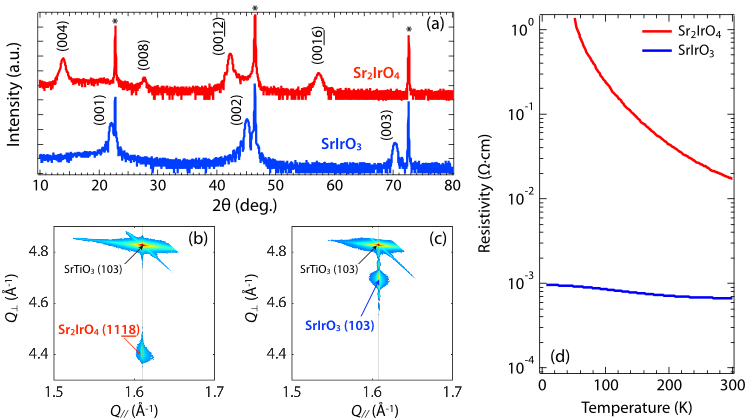}
\caption{\label{} (a) XRD 2$\theta$-$\omega$ scans of single phase Sr$_2$IrO$_4$ and SrIrO$_3$ thin films with thickness $\sim$ 20 nm. The sharp starred peaks correspond to the ($00l$) reflections from the SrTiO$_3$ substrate. (b)-(c) RSM around the SrTiO$_3$ (103) reflection of Sr$_2$IrO$_4$ (b) and SrIrO$_3$ (c). The reflections with different index from the films are labeled on the figures. (d) Temperature dependent resistivity curves of single phase Sr$_2$IrO$_4$ and SrIrO$_3$ thin films.}
\end{figure*}

The spin-orbit doublet of sample (i) at $\sim$ 60.8 eV for Ir 4$f_{7/2}$ and $\sim$ 63.9 eV for Ir 4$f_{5/2}$ correspond to Ir$^0$ valence state \cite{Shan_ChemMater_2007, Kwon_AFM_2017}, indicating that the film is mostly composed of clusters of unoxidized Ir metals. However, the spectra of sample (ii) -- (iv) all exhibit doublet peaks at $\sim$ 62.1 eV for Ir 4$f_{7/2}$ and $\sim$ 65.2 eV for Ir 4$f_{5/2}$, confirming the formation of Ir$^{4+}$ as a result of oxidation by the excess oxygen background, which is consistent with the observations from Sr 3$d$ spectra. In addition, the resultant Sr/Ir intensity ratio is calculated to be $\sim$ 1.8 and 1.2 for Sr214 and Sr113, respectively. The small deviation from the ideal stoichiometric ratio is likely due to the error induced during background subtraction. This ratio confirms the expected chemical composition of each film. Surprisingly, the Sr/Ir ratio for the samples with 3D growth is found to be $\sim$ 4.5 for sample (i) and $\sim$ 3.5 for sample (iii) implying a substantial excess of Sr in those samples.   



In order to link the observed unusual phenomena to the underlying growth mechanism, we consider several distinct stages of the deposition process. First, we note that because of a high laser fluence of $\sim$ 2 J/cm$^2$, the preferential ablation of various species  is likely eliminated \cite{Nishio_APL Mater_2016}. Therefore we can attribute the issues of stoichiometry and growth mode to the next two stages of the PLD process, namely, to the dynamics of plume propagation and the initial stage of nucleation on the substrate surface. It has been generally recognized that during the PLD growth of multicomponent oxide thin films, the background gas pressure can induce significant variations of the angular distribution of different species in the plume, which in turn results in  off-stoichiometry of films with respect to the chemical composition of a target. \cite{Sambri_JAP_2016,Konomi_JVSTA_2008,Droubay_APL_2010,Canulescu_JAP_2009,Amoruso_ASS_2012} More specifically, in nearly vacuum, the plume is primarily composed of elemental atomic/ionic species. Compared to heavier ions,  lighter species tend to propagate with higher velocity and hence a narrower angular distribution towards the substrate normal.\cite{Sambri_JAP_2016, Konomi_JVSTA_2008} This corresponds to the case of sample ($i$), where the concentration of the lighter Sr related species exceeds the concentration of  Ir species at the forefront part of the plume, thus resulting in a Sr enriched phase on the substrate. Furthermore, this regime results in a 3D growth mode due to the reduced surface mobility of the adatoms compared to the large kinetic energy of the subsequently  arriving  species from the plume. The increasing amount of  $P_{O_2}$ allows the plume species to be gradually oxidized and broadens the plume angular distribution. \cite{Sambri_JAP_2016,Konomi_JVSTA_2008,Droubay_APL_2010,Canulescu_JAP_2009,Amoruso_ASS_2012,Sambri_JAP_2008,Amoruso_JAP_2010}. In particular, under an intermediate background pressure (sample ($ii$), $P_{O_2} \sim$ 10$^{-3}$ Torr), this broadening effect for lighter species is more pronounced compared  to the heavier ones. As a result, the stoichiometry of the plume front becomes commensurate with that of target.\cite{Sambri_JAP_2016} 
Furthermore, under high background pressure  (sample ($iv$) $P_{O_2} \sim 10^{-1}$ Torr), lighter species are likely to experience back-scattering by background oxygen. \cite{Sambri_JAP_2016,Wicklein_APL_2012,Ohnishi_APE_2011,Groenendijk_APL_2016}; Eventually a crossover to  the regime  where the angular distribution of lighter species is broader than that of the heavier ones takes place. This mechanism gives rise to the formation of a Ir-rich phase in the film. In each described case, the adatoms gain sufficient mobility determined primarily by $T_{s}$ to propagate and nucleate before the arrival  of subsequent species; this condition leads to the desired 2D or layer-by-layer growth mode. It is worth noting that if $P_{O_2}$ is set at 10$^{-3}$ Torr and $T_{s}$  reduces down to 600 $^{\circ}$C (i.e. moving from sample (ii) to sample (iii)), the Sr-enriched phase emerges again. Interestingly, similar phenomena were also observed during the PLD growth of perovskite manganite and titanite thin films \cite{Sambri_JAP_2008,Sambri_JAP_2016,Sambri_APL_2007}, where it was found that $T_{s}$ can effectively tune the shape and concentration of species within the plume. Specifically lowering $T_{s}$ will make the lighter species redistribute preferentially toward the plume front with faster velocity, while the heavier ones largely remain at the tail. In analogy, the distribution of Sr and Ir species in this case is similar to that of (i), leading to the formation of a Sr-enriched phase. These analysis are depicted as the insets of Fig.~3.

Fig.~4(a) presents the high resolution XRD 2$\theta$-$\omega$ scans for both Sr214 and Sr113 thin films deposited under the optimized conditions specifically for each material. The exclusive presence of $(00\underline{4l})$ peaks for Sr214 \cite{Shibuya_APL_2008} and $(00l)$ peaks for Sr113 unambiguously confirms that single crystal phases of Sr214 and Sr113  are indeed stabilized without any secondary chemical phases within the resolution of XRD. Both films are about $\sim$ 20 nm thin, as evidenced from the interval between consecutive Kessig fringes around the film peaks. The extracted from the data out-of-plane lattice parameter $c$ is 4.01 \AA ~for Sr113 (under compressive strain of $\sim$ -1.26\%) and 25.63 \AA ~for Sr214 (under tensile strain of $\sim$ +0.52\%), respectively. These values deviate only slightly from their bulk values given in Table I and are perfectly consistent with the expected strain effect induced by the epitaxy with STO substrate. To investigate the epitaxial relationship and strain, reciprocal  space mapping (RSM) was performed for both samples. As shown in Fig.s ~4(b) and 4(c), the RSMs around the STO asymmetric (103) reflections corroborate with the materials phase identified for each film: namely, the presence of the (103) reflection for the Sr113 film confirms the perovskite structure, while the observed (11\underline{18}) reflection for Sr214 provides a strong evidence for the layered perovskite K$_2$NiF$_4$-type structure with $c$-axis oriented out of the sample surface \cite{Nichlos_APL_2013}. Moreover, both films are coherently strained to the substrate with no detectable strain relaxation as evidenced by the identical with STO substrate in-plane $Q$ values. As the result, the axial ratios $c/a=1.03$ for Sr113 and 3.28 for Sr214 are very close to the reported results grown with separate stoichiometric targets \cite{Jian_arkiv_2013, Biswas_JAP_2014, Jian_PRB_2013}.

\subsection{Electronic properties}

Next we turn our attention to the electronic properties of each single-phase sample. The temperature-dependent resistivity of both samples was measured from 300 K down to 5 K, and is displayed in Fig.~4(d). In close correspondence with the recently reported results for Sr$_{n+1}$Ir$_n$O$_{3n+1}$ epitaxial thin films, the Sr214 sample exhibits a strongly insulating behavior down to the base temperature, whereas the overall resistivity of Sr113 sample barely changes and remains rather low ($\sim$ 800 $\mu\Omega$ $\cdot$ cm) in the whole temperature range indicative of a semi-metallic behavior.\cite{Hirai_APLMater_2015} The estimated activation band gap at 300 K is approximately $\sim$ 120 meV for Sr214, smaller than the bulk value $\sim$ 200 meV \cite{Ge_PRB_2011} due to the strain effect \cite{Jian_PRB_2013}.

\begin{figure}[t]\vspace{-0pt}
\center
\includegraphics[width=0.4\textwidth]{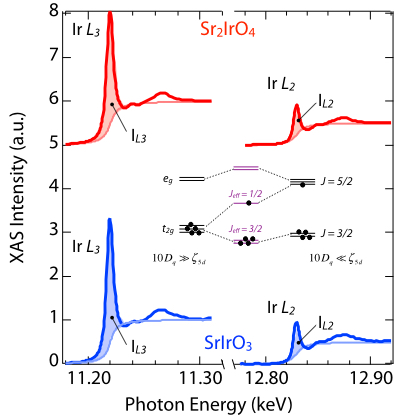}
\caption{\label{} XAS on the Ir $L_{2,3}$ edges of single phase Sr214 and Sr113 thin films. The areas shaded in color represent the white line intensities of each peak by removing the edge-jump background. The inset shows a sketch of the orbital splitting of Ir 5$d$ levels under strong crystal field effect ($10D_q \gg \zeta_{5d}$) and strong SOC ($10D_q \ll \zeta_{5d}$).}
\end{figure}

Given the proposed importance of the SOC effect for the layered iridate compounds, we carried out the XAS measurements on  Ir $L_{2,3}$ edge shown in Fig.~5. We recall that the branching ratio $BR = I_{L_3}/I_{L_2}$ is a  measure of the SOC and represents the integrated white line intensity at each absorption edge. The obtained $BR$ values are 4.9(7) for Sr214, and 4.6(8) for Sr113. Note, those numbers are more than twice as much as the statistical $BR \sim$ 2 indicating the presence of a very strong SO interaction in the iridate thin films.\cite{Clancy_PRB_2012} Moreover, according to the branching ratio analysis, the expectation value of the SOC operator $\expval{\vec{L}\cdot\vec{S}}$ is directly related to the value of $BR$ via $BR = (2+r)/(1-r)$, where $r = \expval{\vec{L}\cdot\vec{S}} / \expval{n_h}$ and $\expval{n_h}$ refers to the number of 5$d$ holes.\cite{Thole_PRL_1988} Taking $\expval{n_h} \approx$ 5 for both films yeilds $\expval{\vec{L}\cdot\vec{S}} =2.4(9)$ for Sr214 and 2.3(6) for Sr113. These values agree very well with the reported result for the bulk compounds \cite{Haskel_PRL_2012, Marco_PRL_2010} thus implying the formation of the spin-orbit entangled $J_\textrm{eff} = 1/2$ electronic ground state(see inset in Fig.~5).      

\begin{figure}[t]\vspace{-0pt}
\center
\includegraphics[width=0.45\textwidth]{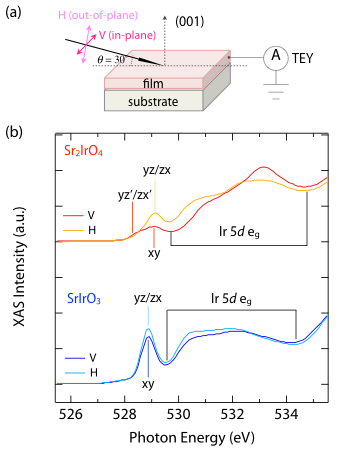}
\caption{\label{} (a) Schematic of the experimental setup. The spectra were recorded separately using horizontal (H) and vertical (V) polarized X-rays. Data was collected in the TEY detection mode. (b) O $K$-edge polarization dependent XAS from single phase thin films of Sr214 (top) and Sr113 (bottom). Note, the yz/zx and xy represent the transition to in-plane oxygens, while yz'/zx' for the transition to apical oxygens if any.}
\end{figure}

In addition, more information about the Ir 5$d$ -- O 2$p$ hybridization can be obtained by measuring the polarization dependent XAS at the O $K$-edge of each single phase film. The relative geometry of the polarization vector $\vec{E}$ with respect to the film orientation is displayed in Fig.~6(a). In general, with linearly polarized X-rays, one can search for the valence holes, and the polarization dependence of the XAS intensity is primarily determined by the orientation of the hole-carrying orbitals. In our case, by selectively probing the holes on different oxygen 2$p$ orbitals (i.e. $p_x$, $p_y$, $p_z$) that are strongly hybridized with Ir 5$d$ states, we can deduce the distribution of holes on the Ir 5$d$ orbitals. As previously shown for the layered perovskite compounds \cite{Mizokawa_PRL_2001}, due to the local tetrahedral distortion, the Ir $d_{xy}$ and $d_{x^2-y^2}$ orbitals can only hybridize with $p_{x,y}$ orbitals from the four in-plane oxygens while the Ir $d_{z^2}$ orbitals can only hybridize with $p_z$ orbitals from the two apical oxygens. The remaining $d_{xz/yz}$ orbitals are allowed to hybridize with both apical oxygen $p_{x/y}$ and in-plane oxygen $p_z$ orbitals.\cite{Mizokawa_PRL_2001,Noh_PRB_2005} Following this scenario, the weak peak at $\sim$ 528.3 eV of Sr214 in Fig.~6(b) is a signature of the apical O $p_{x/y}$ -- Ir $d_{xz/yz}$ bonding state, as is only probed with in-plane polarized X-rays. The much more profound peak at $\sim$ 529.1 eV represents the bondings between in-plane O $p$ and Ir $t_{2g}$ . In particular, the XAS intensity from out-of-plane polarization is about twice as large as that from in-plane polarization, which is a hallmark of the equal orbital population ($d_{xy}:d_{yz}:d_{zx} = 1:1:1$) contributed to the $J_\textrm{eff}=1/2$ ground state.\cite{Kim_PRL_2008} Furthermore, the features between 530 eV and 535 eV are due to the transitions from O $1s \rightarrow 2p$ levels that are hybridized with Ir 5$d$ $e_g$ states. On the other hand, for Sr113 only one distinct peak appears at $\sim$ 528.8 eV in the pre-edge region with the broad Ir 5$d$ $e_g$ -- O $p$ states seen between 529.5 eV and 534.5 eV. Compared to Sr214, the tetragonal distortion is significantly suppressed in Sr113, effectively lowering the $d_{xy}$ energy levels and intermixing the J$_\textrm{eff}=1/2$ with J$_\textrm{eff}=3/2$ states, so that the distribution of holes at each orbital composition becomes $d_{xy}:d_{yz}/d_{zx} < 1:1$.\cite{Kim_JPCM_2014} Again, as shown in Fig.~6(b) for Sr113 sample this observation is consistent with the slightly larger XAS intensity from out-of-plane polarization than that from in-plane.

\section{Conclusion}

In summary, we have demonstrated the selective synthesis of high-quality single phase Sr214 and Sr113 epitaxial thin films of the RP series from a single polycrystalline Sr$_2$IrO$_4$ target. Additionally, we revealed that the oxygen partial pressure plays a decisive role in the cation stoichiometric ratio of the film, while the substrate temperature must be manipulated accordingly to maintain the 2D growth mode. Branching ratio analysis of the XAS results at Ir $L$-edge confirm the large SOC in both films. Polarization dependence of the absorption spectra at O $K$-edge further elucidate the Ir 5$d$ -- O 2$p$ hybridization in each single phase film. Our findings provide valuable insight into the growth mechanism of Sr$_{n+1}$Ir$_n$O$_{3n+1}$ films at non-equilibrium conditions, and pave another way for the fabrications of a variety of RP phases by kinetic control.

\section{Acknowledgements}

The authors deeply acknowledge Jaewook Kim and Mike Hawkridge for their assistance on electrical transport and X-ray diffraction measurements, respectively. X.L. also thank Heung-sik Kim for fruitful discussions. X.L., Y.C. and J.C. were supported by the Gordon and Betty Moore Foundation's EPiQS Initiative through Grant GBMF4534, and by the Department of Energy under grant DE-SC0012375. This research used resources of the Advanced Light Source, which is a Department of Energy Office of Science User Facility under Contract No. DE-AC0205CH11231. This research used resources of the Advanced Photon Source, a U.S. Department of Energy Office of Science User Facility operated by Argonne National Laboratory under Contract No. DE-AC02-06CH11357.


\begin{thebibliography}{999}

\bibitem{Krempa_ARCMP_2014}
W. Witczak-Krempa, G. Chen, Y. Kim, and L. Balents, Annu. Rev. Condens. Matter Phys. \textbf{5}, 57-82 (2014).

\bibitem{Schaffer_RPP_2016}
R. Schaffer, E. Lee, B. Yang, and Y. Kim, Rep. Prog. Phys. \textbf{79}, 094504 (2016).

\bibitem{Moon_PRL_2008}
S. J. Moon \textit{et al.}, Phys. Rev. Lett. \textbf{101}, 226402 (2008).

\bibitem{Kim_PRL_2008}
B. J. Kim \textit{et al.}, Phys. Rev. Lett. \textbf{101}, 076402 (2008).

\bibitem{Kim_Science_2009}
B. J. Kim, H. Ohsumi, T. Komesu, S. Sakai, T. Morita, H. Takagi, T. Arima, Science \textbf{323}, 1329 (2009).

\bibitem{Kim_PRL_2012}
J. Kim \textit{et al.}, Phys. Rev. Lett. \textbf{108}, 177003 (2012).

\bibitem{Kim_NPhys_2015}
Y. K. Kim, N. H. Sung, J. D. Denlinger, and B. J. Kim, Nat. Phys. \textbf{12}, 37-41 (2016).

\bibitem{Zhao_NPhys_2015}
L. Zhao, D. H. Torchinsky, H. Chu, V. Ivanov, R. Lifshitz, R. Flint, T. Qi, G. Cao, and D. Hsieh, Nat. Phys. \textbf{12}, 32-36 (2016).

\bibitem{Yan_PRX_2015}
Y. J. Yan \textit{et al.}, Phys. Rev. X \textbf{5}, 041018 (2015).

\bibitem{Wang_PRL_2011}
F. Wang and T. Senthil, Phys. Rev. Lett. \textbf{106}, 136402 (2011).

\bibitem{Watanabe_PRL_2013}
H. Watanabe, T. Shirakawa, and S. Yunoki, Phys. Rev. Lett. \textbf{110}, 027002 (2013).

\bibitem{Meng_PRL_2014}
Z. Y. Meng, Y. B. Kim, and H.-Y. Kee, Phys. Rev. Lett. \textbf{113}, 177003 (2014).

\bibitem{Zeb_PRB_2012}
M. Ahsan Zeb, and H. Kee, Phys. Rev. B \textbf{86}, 085149 (2012).

\bibitem{Xiao_NC_2011}
D. Xiao \textit{et al.}, Nature Commun. \textbf{2}, 596 (2011).

\bibitem{Carter_PRB_2012}
J.-M. Carter, V. Vijay Shankar, M. Ahsan Zeb, and H.-Y Kee, Phys. Rev. B \textbf{85}, 115105 (2012).

\bibitem{Chen_NC_2015}
Y. Chen, Y.-M. Lu, and H.-Y. Kee, Nature Commun. \textbf{6}, 6593 (2015).

\bibitem{Jian_arkiv_2013}
J. Liu \textit{et al.}, arXiv:1305.1732v1.

\bibitem{Lee_PRB_2012}
J. S. Lee, Y. Krockenberger, K. S. Takahashi, M. Kawasaki, and Y. Tokura, Phys. Rev. B \textbf{85}, 035101 (2012).

\bibitem{Jian_PRB_2013}
C. Rayan Serrao \textit{et al.}, Phys. Rev. B \textbf{87}, 085121 (2013).

\bibitem{Nichlos_APL_2013}
J. Nichlos, J. Terzic, E. G. Bittle, O. B. Korneta, L. E. De Long, J. W. Brill, G. Cao, and S. S. A. Seo, Appl. Phys. Lett. \textbf{102}, 141908 (2013).

\bibitem{Nichlos_APL_2013_2}
J. Nichlos, O. B. Korneta, J. Terzic, L. E. De Long, G. Cao, J. W. Brill, and S. S. A. Seo, Appl. Phys. Lett. \textbf{103}, 131910 (2013).

\bibitem{Miao_PRB_2014}
L. Miao, H. Xu, and Z. Q. Mao, Phys. Rev. B \textbf{89}, 035109 (2014).

\bibitem{Biswas_JAP_2014}
A. Biswas, K.-S. Kim, and Y. H. Jeong, J. Appl. Phys. \textbf{116}, 213704 (2014).

\bibitem{Zhang_PRB_2015}
L. Zhang \textit{et al.}, Phys. Rev. B \textbf{91}, 035110 (2015).

\bibitem{Nie_PRL_2015}
Y. F. Nie \textit{et al.}, Phys. Rev. Lett. \textbf{114}, 016401 (2015).

\bibitem{Liu_Srep_2016}
Z. T. Liu \textit{et al.}, Sci. Rep. \textbf{6}, 30309 (2016).

\bibitem{Fina_NC_2013}
I. Fina \textit{et al.}, Nature Commun. \textbf{5}, 4671 (2013).

\bibitem{Matsuno_PRL_2015}
J. Matsuno, K. Ihara, S. Yamamura, H. Wadati, K. Ishii, V. V. Shankar, H.-Y. Kee, and H. Takagi, Phys. Rev. Lett. \textbf{114}, 247209 (2015).

\bibitem{Groenendijk_APL_2016}
D. J. Groenendijk, N. Manca, G. Mattoni, L. Kootstra, S. Gariglio, Y. Huang, E. van Heumen, and A. D. Caviglia, Appl. Phys. Lett. \textbf{109}, 041906 (2016).

\bibitem{Hirai_APLMater_2015}
D. Hirai. J. Matsuno, and H. Takagi, APL Mater. \textbf{3}, 041508 (2015).

\bibitem{Yi_PNAS_2015}
D. Yi \textit{et al.}, PNAS. \textbf{113}, 6397-6402 (2016).

\bibitem{Nichols_NC_2016}
J. Nichols \textit{et al.}, Nature Commun. \textbf{7}, 12721 (2016).

\bibitem{Nishio_APL Mater_2016}
K. Nishio, H. Y. Hwang, and Y. Hikita, APL Mater. \textbf{4}, 036102 (2016).

\bibitem{Seo_APL_2016}
S. S. A. Seo, J. Nichoos, J. Hwang, J. Terzic, J. H. Gruenewald, M. Souri, J. Thompson, J. G. Connell, and G. Cao, Appl. Phys. Lett. \textbf{109}, 201901 (2016).

\bibitem{Gruenewald_AM_2017}
J. H. Gruenewald \textit{et al.}, Adv. Mater. \textbf{29}, 1603797 (2017).

\bibitem{Hao_PRL_2017}
L. Hao \textit{et al.}, Phys. Rev. Lett. \textbf{119}, 027204 (2017).

\bibitem{Okamoto_NanoLett_2017}
S. Okamoto, J. Nichols, C. Sohn, S. Y. Kim, T. W. Noh, and H. N. Lee, Nano Lett. \textbf{17}, 2126-2130 (2017).

\bibitem{Longo_JSSC_1971}
J. M. Longo, J. A. Kafalas, and R. J. Arnott, J. Solid State Chem. \textbf{3}, 174-179 (1971).

\bibitem{SI}
See Supplemental Material.

\bibitem{Ichimiya_RHEED}
A. Ichimiya and P. I. Cohen, \textit{Reflection High-Energy Electron Diffraction} (Cambridge University Press, Cambridge, 2004). 

\bibitem{Mutoro_EES_2011}
E. Mutoro, E. J. Crumlin, M. D. Biegalski, H. M. Christen, and Y. Shao-Horn, Energy Environ. Sci. \textbf{4}, 3689 (2011). 

\bibitem{Xu_SciRep_2016}
C. Xu, H. Du, A. J. H. van der Torren, J. Aarts, C.-L. Jia, and R. Dittmann, Sci. Rep. \textbf{6}, 38296 (2016).

\bibitem{Yang_SciRep_2017}
W. C. Yang \textit{et al.}, Sci. Rep. \textbf{7}, 7740 (2017).

\bibitem{Pfeifer_SIA_2015}
V. Pfeifer \textit{et al.}, Surf. Interface Anal. \textbf{48}, 261-273 (2016).

\bibitem{Kahk_PRL_2014}
J. M. Kahk \textit{et al.}, Phys. Rev. Lett. \textbf{112}, 117601 (2014).

\bibitem{Payne_CPL_2005}
D. J. Payne, R. G. Egdell, W. Hao, J. S. Foord, A. Walsh, and G. W. Watson, Chem. Phys. Lett. \textbf{411}, 181-185 (2005).

\bibitem{Bourlange_JAP_2009}
A. Bourlange \textit{et al.}, J. Appl. Phys. \textbf{106}, 013703 (2009).

\bibitem{Kotani_JESRP_1996}
A. Kotani, J. Electron Spectrosc. Relat. Phenom. \textbf{78}, 7 (1996).

\bibitem{Shan_ChemMater_2007}
C. Shan, D. Tsai, Y. Huang, S. Jian, and C. Cheng, Chem. Mater. \textbf{19}, 424-431 (2007).

\bibitem{Kwon_AFM_2017}
T. Kwon, H. Hwang, Y. Sa, J. Park, H. Baik, S. Joo, and K. Lee, Adv. Funct. Mater. \textbf{27}, 1604688 (2017).



\bibitem{Sambri_JAP_2016}
A. Sambri, C. Aruta, E. Di Gennaro, X. Wang, U. Scotti di Uccio, F. Miletto Granozio, and S. Amoruso, J. Appl. Phys. \textbf{119}, 125301 (2016).

\bibitem{Konomi_JVSTA_2008}
I. Konomi, T. Motohiro, M. Horii, and M. Kawasumi, J. Vac. Sci. Technol. A \textbf{26}, 1455 (2008).

\bibitem{Droubay_APL_2010}
T. C. Droubay, L. Qiao, T. C. Kaspar, M. H. Engelhard, V. Shutthanand, and S. A. Chambers, Appl. Phys. Lett. \textbf{97}, 124105 (2010).

\bibitem{Canulescu_JAP_2009}
S. Canulescu, E. L. Papadopoulou, D. Anglos, Th. Lippert, C. W. Schneider, and A. Wokaun, J. Appl. Phys. \textbf{105}, 063107 (2009).

\bibitem{Amoruso_ASS_2012}
S. Amoruso, C. Aruta, P. Aurino, R. Bruzzese, X. Wang, F. Miletto Granozio, and U. Scotti di Uccio, Appl. Surf. Sci. \textbf{258}, 9116-9122 (2012).



\bibitem{Sambri_JAP_2008}
A. Sambri, S. Amoruso, X. Wang, F. Miletto Granozio, and R. Bruzzese, J. Appl. Phys. \textbf{104}, 053304 (2008).

\bibitem{Amoruso_JAP_2010}
S. Amoruso, C. Aruta, R. Bruzzese, D. Maccariello, L. Maritato, F. Miletto Granozio, P. Orgiani, U. Scotti di Uccio, and X. Wang, J. Appl. Phys. \textbf{108}, 043302 (2010).

\bibitem{Wicklein_APL_2012}
S. Wicklein, A. Sambri, S. Amoruso, X. Wang, R. Bruzzese, A. Koehl, and R. Dittmann, Appl. Phys. Lett. \textbf{101}, 131601 (2012).

\bibitem{Ohnishi_APE_2011}
T. Ohnishi, and K. Takada, Appl. Phys. Express \textbf{4}, 025501 (2011).

\bibitem{Sambri_APL_2007}
A. Sambri, S. Amoruso, X. Wang, M. Radovic, F. Miletto Granozio, and R. Bruzzese, Appl. Phys. Lett. \textbf{91}, 151501 (2007).


\bibitem{Shibuya_APL_2008}
K. Shibuya, S. Mi, C.-L. Jia, P. Meuffels, and R. Dittmann, Appl. Phys. Lett. \textbf{92}, 241918 (2008).

\bibitem{Ge_PRB_2011}
M. Ge, T. F. Qi, O. B. Korneta, D. E. De Long, P. Schlottmann, W. P. Crummett, and G. Cao, Phys. Rev. B \textbf{84}, 100402(R) (2011).


\bibitem{Clancy_PRB_2012}
J. P. Clancy, N. Chen, C. Y. Kim, W. F. Chen, K. W. Plumb, B. C. Jeon, T. W. Noh, and Y.-J. Kim, Phys. Rev. B \textbf{86}, 195131 (2012).

\bibitem{Thole_PRL_1988}
G. van der Laan, and B. T. Thole, Phys. Rev. Lett. \textbf{60}, 1977 (1988).

\bibitem{Haskel_PRL_2012}
D. Haskel, G. Fabbris, M. Zhernenkov, P. P. Kong, C. Q. Jin, G. Cao, and M. van Veenendaal, Phys. Rev. Lett. \textbf{109}, 027204 (2012).

\bibitem{Marco_PRL_2010}
M. A. Laguna-Marco, D. Haskel, N. Souza-Neto, J. C. Lang, V. V. Krishnamurthy, S. Chikara, G. Cao, and M. van Veenendaal, Phys. Rev. Lett. \textbf(105), 216407 (2010).

\bibitem{Mizokawa_PRL_2001}
T. Mizokawa, L. H. Tjeng, G. A. Kawatzky, G. Ghiringhelli, O. Tjernberg, N. B. Brookes, H. Fukazawa, S. Nakatsuji, and Y. Maeno, Phys. Rev. Lett. \textbf{87}, 077202 (2001).

\bibitem{Noh_PRB_2005}
H.-J. Noh \textit{et al.}, Phys. Rev. B \textbf{72}, 052411 (2005).

\bibitem{Kim_JPCM_2014}
K.-H. Kim, H.-S. Kim, and M. J. Han, J. Phys.: Condens. Matter \textbf{26}, 185501 (2014).




\end{thebibliography}
\end{document}